\documentclass[12pt,a4paper]{article}
\usepackage[dvips]{graphicx}
\usepackage{amsmath}
\usepackage{amsfonts}
\usepackage{amssymb}

\begin{document}

\author{S. Gov and S. Shtrikman\thanks{Also with the Department of Physics, University
of California, San Diego, La Jolla, 92093 CA, USA.}\\The Department of Electronics, \\Weizmann Institute of Science,\\Rehovot 76100, Israel
\and H. Thomas\\The Department of Physics and Astronomy,\\University of Basel,\\CH-4056 Basel, Switzerland}
\title{1D Toy Model For Trapping Neutral Particles}
\date{}
\maketitle
\begin{abstract}
We study, both classically and quantum-mechanically, the problem of a neutral
particle with a spin $S$, mass $m$ and magnetic moment $\mu$, moving in one
dimension in an inhomogeneous magnetic field given by
\[
\mathbf{B}=B_{0}\mathbf{\hat{z}+}B_{\bot}^{\prime}x\mathbf{\hat{y}.}%
\]
This problem serves for us as a toy model to study the trapping of neutral
particles. We identify
\[
K\equiv\sqrt{\dfrac{S^{2}\left(  B_{\bot}^{\prime}\right)  ^{2}}{\mu
mB_{0}^{3}}},
\]
which is the ratio between the precessional frequency of the particle and its
vibration frequency, as the relevant parameter of the problem.

Classically, we find that when $\mu$ is antiparallel to $\mathbf{B}$, the
particle is trapped provided that $K<0.5$. We also find that viscous friction,
be it translational or precessional, destabilizes the system.

Quantum-mechanically, we study the problem of a spin $S=\hbar/2$ particle in
the same field. Treating $K$ as a small parameter for the perturbation from
the adiabatic Hamiltonian, we find that the lifetime $T_{esc}$ of the particle
in its trapped ground-state is
\[
T_{esc}=T_{vib}\sqrt{\dfrac{1}{\left(  2\pi\right)  ^{3}K}}\exp\left[
\dfrac{1}{2K}\right]  \text{ ,}%
\]
where $T_{vib}=2\pi\sqrt{mB_{0}/\mu\left(  B_{\bot}^{\prime}\right)  ^{2}}$ is
the classical period of the particle when placed in the adiabatic potential
$V=\mu\left|  \mathbf{B}\right|  $.
\end{abstract}

\section{Introduction.\label{intro}}

\subsection{Magnetic traps for neutral particles.\label{traps}}

Recently there has been rapid progress in techniques for trapping samples of
neutral atoms at elevated densities and extremely low temperatures. The
development of magnetic and optical traps for atoms has proceeded in parallel
in recent years. While optical methods have proved to be an efficient means of
cooling atoms to temperatures in the microKelvin range, further progress is
limited by interatomic interactions induced by the scattering of photons. The
effort to attain higher densities and lower temperatures has therefore
concentrated on the development of purely magnetic
traps\cite{t1,t2,t3,t4,t5,t6,bec}. Such traps exploit the interaction of the
magnetic moment of the atom with the inhomogeneous magnetic field to provide
spatial confinement.

Microscopic particles are not the only candidates for magnetic traps. In fact,
a vivid demonstration of trapping large-scale objects is the hovering magnetic
top\cite{levitron,patent,ucas}. This ingenious magnetic device, which hovers
in mid-air for about 2 minutes, has been studied recently by several authors
\cite{Berry,bounds,simon,dynamic}.

\subsection{Qualitative description.\label{desc}}

The physical mechanism underlying the operation of magnetic traps is the
adiabatic principle. The common way to describe their operation is in terms of
\emph{classical} mechanics: As the particle is released into the trap, its
magnetic moment points antiparallel to the direction of the magnetic field.
While inside the trap, the particle experiences lateral oscillations
$\omega_{vib}$ which are slow compared to its precession $\omega_{prec}$.
Under this condition the spin of the particle may be considered as
experiencing a \emph{slowly} rotating magnetic field. Thus, the spin precesses
around the \emph{local} direction of the magnetic field $\mathbf{B}$
(adiabatic approximation) and, on the average, its magnetic moment
$\mathbf{\mu}$ points \emph{antiparallel} to the local magnetic field lines.
Hence, the magnetic energy, which is normally given by $-\mathbf{\mu}%
\cdot\mathbf{B}$, is now given (for small precession angle) by $\mu\left|
\mathbf{B}\right|  $. Thus, the overall effective potential seen by the
particle is
\begin{equation}
V_{eff}\simeq\mu\left|  \mathbf{B}\right|  .\label{energy}%
\end{equation}
In the adiabatic approximation, the spin degree of freedom is rigidly coupled
to the translational degrees of freedom, and is already incorporated in
Eq.(\ref{energy}). Thus, under the adiabatic approximation, the particle may
be considered as having only translational degrees of freedom. When the
strength of the magnetic field possesses a\emph{\ minimum}, the effective
potential becomes attractive near that minimum and the whole apparatus acts as
a trap. To prevent spin-flip (Majorana transitions), most magneto-static traps
include a bias field, so that the effective potential $V_{eff}$ possesses a
\emph{nonvanishing} minimum.

As mentioned above, the adiabatic approximation holds whenever $\omega
_{prec}\gg\omega_{vib}$. As $\omega_{prec}$ is inversely proportional to the
spin, this inequality can be satisfied provided that the spin of the particle
is small enough. If, on the other hand, the spin of the particle is too large,
it cannot respond fast enough to the changes of the direction of the magnetic
field. In this limit $\omega_{prec}\gg\omega_{vib}$, the spin has to be
considered as fixed in space and, according to Earnshaw's
theorem\cite{earnshaw}, becomes unstable against \emph{translations}.

\subsection{The purpose and structure of this paper.\label{purp}}

The discussion of magnetic traps in the literature is, almost entirely, done
in terms of \emph{classical} mechanics. In microscopic systems, however,
quantum effects become dominant, and in these cases \emph{quantum mechanics}
is suited for the description of the trap. An even more interesting issue is
the understanding of how the classical and quantum descriptions of a
\emph{given} system are related. In this paper we study, both classically and
quantum-mechanically, the quantitative nature of magnetic traps. In order to
keep the underlying physics transparent, we devise a simplified model for the
inhomogeneous magnetic field of such traps. We further neglect the effect of
interactions between the particles in the trap and so we analyze the dynamics
of a \emph{single} particle inside the trap. For simplicity, the particle is
considered to have only a single translational degree of freedom. Its spin
degree of freedom, on the other hand, is taken completely into account.

The structure of this paper is as follows: In Sec.\ref{def} we start by
defining the system we study, together with useful parameters that will be
used throughout this paper. Next, we carry out a classical analysis of the
problem in Sec.\ref{class}. Here, we find two stationary solutions for the
particle inside the trap. One of them corresponds to a state whose spin is
\emph{parallel} to the direction of the magnetic field whereas the other one
corresponds to a state whose spin is \emph{antiparallel }to that direction.
When considering the dynamical stability of these solutions, we find that only
the \emph{antiparallel }stationary solution is stable. We also study the same
problem but with viscous friction added, and arrive at the interesting result
that friction \emph{destabilizes} the system. In Sec.\ref{quant} we reconsider
the problem, from a quantum-mechanical point of view. Here, we also find
states that refer to \emph{parallel} and \emph{antiparallel} orientations of
the spin, one of them being bound while the other one unbounded. In this case,
however, these two states are \emph{coupled} due to the inhomogeneity of the
field, and we move on to calculate the \emph{lifetime} of the bound state.
Finally, in Sec.\ref{dis} we compare the results of the classical analysis
with those of the quantum analysis and comment on their implications for
practical magnetic traps.

\section{Description of the problem.\label{def}}

We consider a particle of mass $m$, magnetic moment $\mu$ and intrinsic spin
$S$ (aligned with $\mu$) moving in 1D space in an inhomogeneous magnetic field
$\mathbf{B}$ given by
\begin{equation}
\mathbf{B=}B_{0}\mathbf{\hat{z}+}B_{\bot}^{\prime}x\mathbf{\hat{y}}%
\text{.}\label{d0}%
\end{equation}
This field possesses a nonzero minimum of amplitude at the origin, which is
the essential part of the trap. Note also that the direction of the field
twists ( or curls) as one moves along the $x$ axis. The Hamiltonian for this
system is%
\begin{equation}
H=\dfrac{P^{2}}{2m}-\mathbf{\mu\cdot B}\label{d0.1}%
\end{equation}
where $P$ is the momentum of the particle.

We define $\omega_{prec}$ as the precessional frequency of the particle when
it is at the origin $x=0$. Since at that point the magnetic field is
$\mathbf{B=}B_{0}\mathbf{\hat{z}}$ we find that
\begin{equation}
\omega_{prec}\equiv\dfrac{\mu B_{0}}{S}\text{.}\label{d1}%
\end{equation}
Next, we define $\omega_{vib}$ as the small-amplitude vibrational frequency of
the particle when it is placed in the adiabatic potential field given by
\[
V(x)=\mu\left|  \mathbf{B}(x)\right|  =\mu B_{0}\left(  1+\dfrac{1}{2}\left(
\dfrac{B_{\bot}^{\prime}}{B_{0}}\right)  ^{2}x^{2}\right)  +\mathcal{O}\left(
x^{4}\right)  .
\]
For this potential we have
\[
k=\left.  \dfrac{\partial^{2}V}{\partial x^{2}}\right|  _{x=0}=\mu
\dfrac{\left(  B_{\bot}^{\prime}\right)  ^{2}}{B_{0}}\text{,}%
\]
and therefore
\begin{equation}
\omega_{vib}\equiv\sqrt{\dfrac{k}{m}}=\sqrt{\dfrac{\left(  B_{\bot}^{\prime
}\right)  ^{2}\mu}{mB_{0}}}\text{.}\label{d2}%
\end{equation}
We also define the ratio between $\omega_{vib}$ and $\omega_{prec}$,
\begin{equation}
K\equiv\dfrac{\omega_{vib}}{\omega_{prec}}=\sqrt{\dfrac{S^{2}(B_{\bot}%
^{\prime})^{2}}{\mu mB_{0}^{3}}}\text{.}\label{d3}%
\end{equation}
This will be our `measure of adiabaticity'. It is clear that as $K$ becomes
smaller and smaller, the adiabatic approximation becomes more and more
accurate. Note also that $K$ is the only possibility to form a non-dimensional
quantity (up to an arbitrary power) out of the parameters of the system. The
value of $K$ therefore, completely determines the behavior of the system.

\section{Classical analysis.\label{class}}

\subsection{The stationary solutions.\label{stat}}

We denote by $\mathbf{\hat{n}}$ a unit vector in the direction of the spin
(and the magnetic moment). Thus, the equation of motion for the center of mass
of the particle is
\begin{equation}
m\dfrac{d^{2}x}{dt^{2}}=\mu\dfrac{\partial}{\partial x}\left(  \mathbf{\hat
{n}\cdot B}\right)  \text{,}\label{c1}%
\end{equation}
and the evolution of its spin is determined by
\begin{equation}
S\dfrac{d\mathbf{\hat{n}}}{dt}=\mu\mathbf{\hat{n}\times B}\text{.}\label{c2}%
\end{equation}
The two equilibrium solutions to Eqs.(\ref{c1}) and (\ref{c2}) are
\begin{align}
\mathbf{\hat{n}}(t) &  =\mp\mathbf{\hat{z}}\label{c3}\\
x(t) &  =0\text{,}\nonumber
\end{align}
representing a motionless particle at the origin with its magnetic moment (and
spin) pointing \emph{antiparallel} ($\mathbf{\hat{n}}(t)=-\mathbf{\hat{z}}$)
to the direction of the field at that point and a similar solution but with
the magnetic moment pointing \emph{parallel} to the direction of the field
($\mathbf{\hat{n}}(t)=+\mathbf{\hat{z}}$).

\subsection{Stability of the solutions.}

To check the stability of these solutions we now add first-order
perturbations. We set
\begin{align}
\mathbf{\hat{n}(}t\mathbf{)} &  =\mathbf{\mp\hat{z}+}\epsilon_{x}%
(t)\mathbf{\hat{x}+}\epsilon_{y}(t)\mathbf{\hat{y}}\label{c4}\\
x(t) &  =0+\delta x(t)\text{,}\nonumber
\end{align}
substitute these into Eqs.(\ref{c1}) and (\ref{c2}), and retain only
first-order terms. We find that the resulting equations for $\delta x(t)$,
$\epsilon_{x}(t)$ and $\epsilon_{y}(t)$ are
\begin{align}
\dfrac{d^{2}\delta x}{dt^{2}} &  =\dfrac{\mu B_{\bot}^{\prime}}{m}\epsilon
_{y}\label{c5}\\
\dfrac{d\epsilon_{x}}{dt} &  =\dfrac{\mu}{S}\left(  \epsilon_{y}B_{0}\pm
B_{\bot}^{\prime}\delta x\right)  \nonumber\\
\dfrac{d\epsilon_{y}}{dt} &  =\mathbf{-}\dfrac{\mu}{S}\epsilon_{x}%
B_{0}\text{.}\nonumber
\end{align}
We now look for oscillatory (stable) solutions for these equations. Setting
\begin{align*}
\delta x &  =x_{0}e^{-i\omega t}\\
\epsilon_{x} &  =\epsilon_{x,0}e^{-i\omega t}\\
\epsilon_{y} &  =\epsilon_{y,0}e^{-i\omega t}%
\end{align*}
inside Eqs.(\ref{c5}) yields
\begin{equation}
\underset{\mathbf{A}}{\underbrace{\left(
\begin{array}
[c]{lll}%
\omega^{2} & 0 & \mu B_{\bot}^{\prime}/m\\
\pm\mu B_{\bot}^{\prime}/S & i\omega & \mu B_{0}/S\\
0 & -\mu B_{0}/S & i\omega
\end{array}
\right)  }}\cdot\left(
\begin{array}
[c]{l}%
x_{0}\\
\epsilon_{x,0}\\
\epsilon_{y,0}%
\end{array}
\right)  =\left(
\begin{array}
[c]{l}%
0\\
0\\
0
\end{array}
\right)  \text{.}\label{c8}%
\end{equation}
This equation has non-trivial solutions whenever the determinant of the matrix
$\mathbf{A}$ vanishes. Thus, the secular equation%

\begin{equation}
\dfrac{K^{2}}{\omega_{vib}^{4}}\det\mathbf{A=}-K^{2}\left(  \dfrac{\omega
}{\omega_{vib}}\right)  ^{4}+\left(  \dfrac{\omega}{\omega_{vib}}\right)
^{2}\mp1=0\label{c11}%
\end{equation}
determines the eigenfrequencies $\omega$ of the various possible modes. When
the \emph{lower} sign is taken in Eq.(\ref{c11}), corresponding to a spin
\emph{parallel} to the magnetic field, we find that one of the roots for
$\omega^{2}$ is purely negative. This indicates that one of the roots for
$\omega$ has a positive imaginary part for \emph{any} $K$ and hence, the
solution is unstable. We concentrate now on the other possible solution,
corresponding to $\mathbf{\hat{n}}(t)=-\mathbf{\hat{z}}$: Here, the solutions
for $\omega^{2}$ are given by
\begin{equation}
\omega^{2}=\left(  \frac{1}{2K^{2}}\pm\frac{\sqrt{1-4K^{2}}}{2K^{2}}\right)
\omega_{vib}^{2}.\label{c12}%
\end{equation}
For $K\rightarrow0$, the slow mode (minus sign) represents the vibration of
the particle in the adiabatic potential, and the fast mode (plus sign)
represents the precession of the spin about the $\mathbf{\hat{z}}$ component
of the magnetic field, as is shown explicitly by the form of the eigenvectors.
The general form of these eigenvectors may be written in terms of an arbitrary
amplitude parameter $A$ as%
\begin{equation}
\left(
\begin{array}
[c]{c}%
\delta x_{0}\\
\epsilon_{x,0}\\
\epsilon_{y,0}%
\end{array}
\right)  =\left(
\begin{array}
[c]{c}%
\dfrac{B_{0}}{B_{\perp}^{\prime}}\left(  \dfrac{\omega_{vib}}{\omega}\right)
^{2}\\
-iK\dfrac{\omega}{\omega_{vib}}\\
-1
\end{array}
\right)  A\label{c12.01}%
\end{equation}
which for small $K$ reduces to%
\begin{equation}
\left(
\begin{array}
[c]{c}%
\delta x_{0}\\
\epsilon_{x,0}\\
\epsilon_{y,0}%
\end{array}
\right)  _{vib}=\left(
\begin{array}
[c]{c}%
\dfrac{B_{0}}{B_{\perp}^{\prime}}\\
-iK\\
-1
\end{array}
\right)  A\label{c12.02}%
\end{equation}
for the vibrational mode, and to
\begin{equation}
\left(
\begin{array}
[c]{c}%
\delta x_{0}\\
\epsilon_{x,0}\\
\epsilon_{y,0}%
\end{array}
\right)  _{prec}=\left(
\begin{array}
[c]{c}%
\dfrac{B_{0}}{B_{\perp}^{\prime}}K^{2}\\
-i\\
-1
\end{array}
\right)  A\label{c12.03}%
\end{equation}
for the precessional mode. From Eq.(\ref{c12.02}) we learn that in the
vibrational mode, the amplitudes of the translational motion of the particle
and the $y$-component of its spin are large compared to the amplitude of the
$x$-component of the spin. Furthermore, the ratio $\left(  \delta
x_{0}/\epsilon_{y,0}\right)  _{vib}=-B_{0}/B_{\perp}^{\prime}$ shows that the
amplitudes $\delta x_{0}$ and $\epsilon_{y,0}$ are related in such a manner
that the direction of the spin is \emph{antiparallel} to the direction of the
local magnetic field. Eq.(\ref{c12.03}) for the precessional mode tells us
that the amplitude of the translational motion of the particle is negligible,
thus in this mode the particle is essentially fixed and its spin precesses
around the direction of the magnetic field at the origin.

Due to the coupling between the translational and the precessional degrees of
freedom, the mode frequencies given in Eq.(\ref{c12}), change with increasing
$K$.

A stable solution requires that all mode frequencies $\omega$ be real.
Consequently, stability means that the roots for $\omega^{2}$ are real and
positive. From Eq.(\ref{c12}) we find that this happens when
\begin{equation}
K<0.5.\label{c13}%
\end{equation}
For the following discussion it is also useful to note that in the region
$K<0.5$,
\begin{align}
\omega_{vib}^{2} &  <\omega^{2}<2\omega_{vib}^{2}\text{ ; for the vibrational
mode}\label{c13.1}\\
\omega^{2} &  >2\omega_{vib}^{2}\text{ ; for the precessional mode.}\nonumber
\end{align}

\begin{figure}
[p]
\begin{center}
\fbox{\includegraphics[
height=7.3327in,
width=5.6844in
]%
{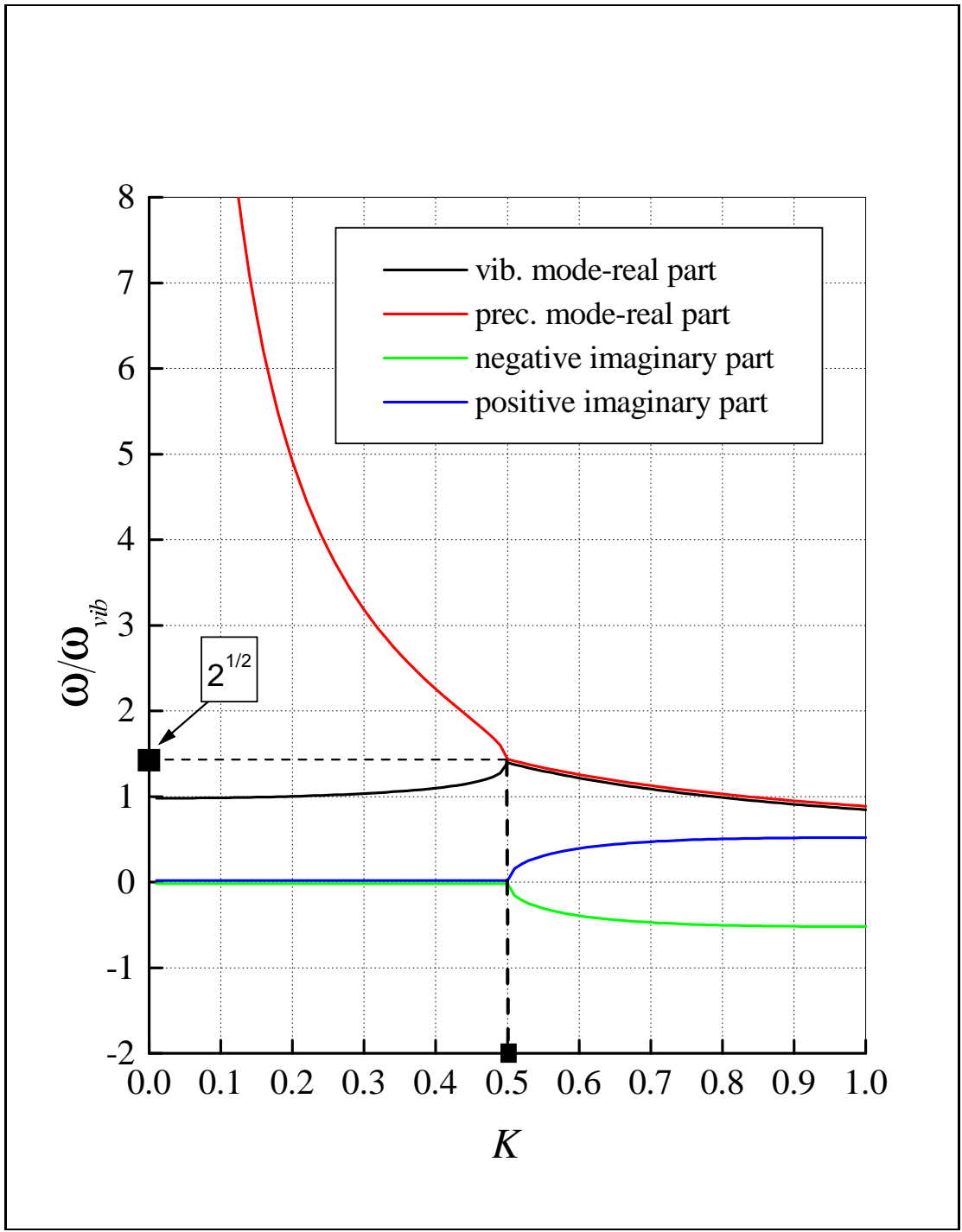}%
}\caption{Real and imaginary parts of the mode frequencies as a function of
$K$.}%
\label{fig1}%
\end{center}
\end{figure}

Fig.\ref{fig1} shows the real and imaginary parts of the frequencies $\omega$
of the two modes as a function of $K$. We note that when $K<0.5$, the
imaginary parts of both frequencies vanish, indicating a stable solution. The
mode with the lower frequency goes asymptotically to $\omega\rightarrow
\omega_{vib}$ when $K\rightarrow0$. This is the \emph{vibrational} mode. The
mode corresponding to the higher frequency, goes asymptotically to
$\omega\rightarrow\omega_{prec}=\omega_{vib}/K$ as $K\rightarrow0$. This is
the \emph{precessional} mode whose frequency, as we mentioned earlier, is
inversely proportional to the spin. When $K>0.5$, both solutions for
$\omega^{2}$ possess an imaginary part and the system becomes unstable.%

\subsection{The excitation energy of the modes.}

The excitation energy of a given mode is defined as the difference between the
energy of the mode and the energy of the stationary state,%
\begin{equation}
e=-\mu\mathbf{\hat{n}}\cdot\mathbf{B+}\dfrac{1}{2}m\left(  \dfrac{d\delta
x}{dt}\right)  ^{2}-\mu B_{0}.\label{ec1.1}%
\end{equation}
Note that the energy is bilinear in the coordinates and hence, one cannot
neglect the $\hat{z}$-component of the spin. Instead, one must set%
\[
\mathbf{\hat{n}\cdot\hat{z}=-}\sqrt{1-\left(  \epsilon_{x}^{2}+\epsilon
_{y}^{2}\right)  }\mathbf{\simeq}-\left(  1-\dfrac{1}{2}\left(  \epsilon
_{x}^{2}+\epsilon_{y}^{2}\right)  \right)  .
\]
Thus, the correct expression of the energy for small amplitudes is%
\begin{equation}
e\simeq-\mu\left(  \dfrac{1}{2}\left(  \epsilon_{x}^{2}+\epsilon_{y}%
^{2}\right)  B_{0}+B_{\perp}^{\prime}\delta x\epsilon_{y}\right)
\mathbf{+}\dfrac{1}{2}m\left(  \dfrac{d\delta x}{dt}\right)  ^{2}%
.\label{ec1.2}%
\end{equation}
In this expression, the modes have to be written in real form,%
\begin{equation}
\delta x\left(  t\right)  =\dfrac{B_{0}}{B_{\perp}^{\prime}}\left(
\dfrac{\omega_{vib}}{\omega}\right)  ^{2}A\cos\left(  \omega t\right)
,\label{ec2}%
\end{equation}%
\begin{equation}
\epsilon_{x}\left(  t\right)  =-K\dfrac{\omega}{\omega_{vib}}A\sin\left(
\omega t\right)  ,\label{ec3}%
\end{equation}%
\begin{equation}
\epsilon_{y}\left(  t\right)  =-A\cos\left(  \omega t\right)  .\label{ec4}%
\end{equation}
Substituting Eqs.(\ref{ec2}), (\ref{ec3}) and (\ref{ec4}) into Eq.(\ref{ec1.2}%
) and using Eq.(\ref{c11}) one finds that%
\begin{equation}
\dfrac{e}{\mu B_{0}}=\dfrac{2\omega_{vib}^{2}-\omega^{2}}{2\omega^{2}}%
A^{2}.\label{ec5}%
\end{equation}
Using Eq.(\ref{c13.1}) we conclude that for $0<K<0.5$, the excitation energy
of the vibrational mode is \emph{positive} while the excitation energy of the
precessional mode is always \emph{negative}. At the point $K=0.5$, where the
two modes coalesce, the excitation energy vanishes. We will further refer to
these observations in the following section.

\subsection{The effect of viscous friction.\label{fric}}

When friction is introduced into the system, the equations of motion become
\begin{equation}
m\dfrac{d^{2}x}{dt^{2}}=\mu\dfrac{\partial}{\partial x}\left(  \mathbf{\hat
{n}\cdot B}\right)  -r_{t}\dfrac{dx}{dt}\label{fric1}%
\end{equation}
and
\begin{equation}
S\dfrac{d\mathbf{\hat{n}}}{dt}=\mu\mathbf{\hat{n}\times B-}r_{p}%
\mathbf{\hat{n}\times}\dfrac{d\mathbf{\hat{n}}}{dt}\text{,}\label{fric2}%
\end{equation}
where $r_{t}$ and $r_{p}$ are translational and precessional friction
coefficients, respectively. The second term on the right-hand side of
Eq.(\ref{fric2}) is the spin-damping contributed by the \emph{change} in the
direction of the spin from $\mathbf{\hat{n}}$ to $\mathbf{\hat{n}%
+}d\mathbf{\hat{n}}$. Since, by definition, $\mathbf{\hat{n}}$ is a unit
vector, $d\mathbf{\hat{n}}$ is perpendicular to $\mathbf{\hat{n}}$. Thus,
$\Omega_{\perp}=\left|  d\mathbf{\hat{n}/}dt\right|  $ is the angular velocity
associated with the change of $\mathbf{\hat{n}}$. Since the direction of
$\Omega_{\perp}$ must be perpendicular to both $d\mathbf{\hat{n}}$ and
$\mathbf{\hat{n}}$ we form the cross product $\mathbf{\Omega}_{\bot
}=\mathbf{\hat{n}\times}\left(  d\mathbf{\hat{n}/}dt\right)  $ which
incorporates both the correct value and the right direction. Multiplying
$\mathbf{\Omega}_{\bot}$ by $r_{p}$ yields the spin-damping term.

To first order in $r_{r}$ and $r_{t}$ the secular equation in this case is
given by$\allowbreak$
\begin{equation}
K^{2}\omega_{n}^{4}-\omega_{n}^{2}-2i\omega_{n}^{3}K\frac{r_{p}}{S}%
+i\frac{\omega_{n}^{3}}{S}\left(  \dfrac{B_{0}}{B_{\perp}^{\prime}}\right)
^{2}r_{t}K^{3}-i\omega_{n}\left(  \dfrac{B_{0}}{B_{\perp}^{\prime}}\right)
^{2}r_{t}\frac{K}{S}+1+i\omega_{n}K\frac{r_{p}}{S}=0\label{fric3}%
\end{equation}
where we defined%
\[
\omega_{n}\equiv\dfrac{\omega}{\omega_{vib}}%
\]
to make the expression simple. Let $\omega_{n,0}$ be the eigenfrequencies
$\omega_{n}$ of the frictionless problem, given by Eq.(\ref{c11}). When adding
a small friction to the problem, the eigenfrequencies will change by a small
amount $\delta\omega_{n}$. We find an approximate expression for $\delta
\omega_{n}$ by expanding Eq.(\ref{fric3}) around $\omega_{n,0}$ to first order
in $\delta\omega_{n}$ and making use of Eq.(\ref{c11}). This gives
\begin{equation}
\delta\omega_{n}=\dfrac{iK}{2S}\left(  \dfrac{r_{p}\omega_{n,0}^{2}\left(
2\omega_{n,0}^{2}-1\right)  +r_{t}\left(  B_{0}/B_{\perp}^{\prime}\right)
^{2}}{\omega_{n,0}^{2}-2}\right)  +\mathcal{O}\left(  r^{2}\right)
.\label{fric4}%
\end{equation}

Eq.(\ref{fric4}) has an interesting consequence: From Eq.(\ref{c13.1}) we find
that the numerator in Eq.(\ref{fric4}) is positive for both modes while the
denominator is negative for the vibrational mode and positive for the
precessional mode. We therefore conclude that friction, either translational
or precessional, stabilizes the vibrational motion and, simultaneously,
destabilizes the precessional motion. The system all together becomes of
course, \emph{unstable}.

The fact that spin damping leads to an exponential growth of the fast mode is
no surprise in view of its negative excitation energy. Also, the exponential
decay of the slow mode due to translational friction is to be expected on
account of its positive excitation energy. What \emph{is} important is the
fact that due to the coupling between translation and precession,
\emph{translational} friction causes an exponential growth of the \emph{fast}
mode, with a growth time which, compared to the effect of spin damping, is
smaller by a factor of $r_{t}K^{2}S^{2}/\mu mr_{p}B_{0}^{2}$ in the limit of
small $K$.

\section{Quantum-mechanical analysis.\label{quant}}

\subsection{The Hamiltonian and its diagonalized form.\label{ham}}

In this section we consider the problem of a neutral particle with spin
$S=\hbar/2$ in a 1D inhomogeneous magnetic field from a quantum-mechanical
point of view. Unlike the classical analysis, in which the derivation was
valid for any value of the adiabaticity parameter $K$, we concentrate here on
the behavior of the system when $K$ is \emph{small}. We choose to analyze the
case of a spin $1/2$\ particle because this case already shows the essentials
of the quantum-mechanical problem.

Now, it is convenient to express the dependence of the magnetic field on $x$
in terms of its amplitude $B(x)$ and its direction $\theta(x)$ with respect to
the $\mathbf{\hat{z}}$ axis. Thus, Eq.(\ref{d0}) is rewritten as
\begin{equation}
\mathbf{B}=B\left(  x\right)  \left(  \sin\left[  \theta\left(  x\right)
\right]  \mathbf{\hat{y}}+\cos\left[  \theta\left(  x\right)  \right]
\mathbf{\hat{z}}\right)  \label{h1}%
\end{equation}
where
\begin{align}
B\left(  x\right)   &  =B\sqrt{1+\left(  \dfrac{B_{\bot}^{\prime}x}{B_{0}%
}\right)  ^{2}}\text{ ,}\label{h1.1}\\
\theta\left(  x\right)   &  =\arctan\left(  \dfrac{B_{\bot}^{\prime}x}{B_{0}%
}\right)  .\nonumber
\end{align}
The time-independent Schr\"{o}dinger equation for this system is
\begin{equation}
\left[  \frac{-\hbar^{2}}{2m}\dfrac{\partial^{2}}{\partial x^{2}}-\mu
B(x)\left(  \sin\left[  \theta\left(  x\right)  \right]  \hat{\sigma}_{y}%
+\cos\left[  \theta\left(  x\right)  \right]  \hat{\sigma}_{z}\right)
\right]  \Psi\left(  x\right)  =E\Psi\left(  x\right)  \label{h5}%
\end{equation}
where $\hat{\sigma}_{y}$ and $\hat{\sigma}_{z}$ are the Pauli matrices given
by
\[%
\begin{array}
[c]{cc}%
\hat{\sigma}_{y}=\left(
\begin{array}
[c]{cc}%
0 & -i\\
i & 0
\end{array}
\right)  \text{ ,} & \hat{\sigma}_{z}=\left(
\begin{array}
[c]{cc}%
1 & 0\\
0 & -1
\end{array}
\right)  ,
\end{array}
\]
$E$ is the eigenenergy, and $\Psi$ is the two-component spinor
\begin{equation}
\Psi=\left(
\begin{array}
[c]{c}%
\psi_{\uparrow}\left(  x\right)  \\
\psi_{\downarrow}\left(  x\right)
\end{array}
\right)  .\label{h5.1}%
\end{equation}
In matrix form Eq.(\ref{h5}) becomes
\begin{equation}
\left(  H_{K}+H_{M}\right)  \left(
\begin{array}
[c]{c}%
\psi_{\uparrow}\left(  x\right)  \\
\psi_{\downarrow}\left(  x\right)
\end{array}
\right)  =E\left(
\begin{array}
[c]{c}%
\psi_{\uparrow}\left(  x\right)  \\
\psi_{\downarrow}\left(  x\right)
\end{array}
\right)  \label{h6.0}%
\end{equation}
where $H_{K}$ and $H_{M}$, given by
\begin{align}
H_{K} &  \equiv\dfrac{-\hbar^{2}}{2m}\left(
\begin{array}
[c]{cc}%
\dfrac{\partial^{2}}{\partial x^{2}} & 0\\
0 & \dfrac{\partial^{2}}{\partial x^{2}}%
\end{array}
\right)  \label{h6.1}\\
H_{M} &  \equiv\mu B\left(  x\right)  \left(
\begin{array}
[c]{cc}%
-\cos\left[  \theta\left(  x\right)  \right]   & i\sin\left[  \theta\left(
x\right)  \right]  \\
-i\sin\left[  \theta\left(  x\right)  \right]   & \cos\left[  \theta\left(
x\right)  \right]
\end{array}
\right)  ,\nonumber
\end{align}
are the kinetic part and the magnetic part of the Hamiltonian $H$, respectively.

In order to diagonalize the magnetic part of the Hamiltonian, we make a local
\emph{passive} transformation of coordinates on the wave function such that
the spinor is expressed in a new coordinate system whose $\mathbf{\hat{z}}$
axis coincides with the direction of the magnetic field at the point $x$. We
denote by $R\left(  x\right)  $ the required transformation and set
$\Psi^{\prime}=R\Psi$. Thus, $\Psi^{\prime}$ represent \emph{the same}
direction of the spin as before the transformation but using the \emph{new}
coordinate system. The Hamiltonian in this newly defined system is clearly
given by $RHR^{-1}$. In the case of the magnetic field given in Eq.(\ref{h1}),
the required operation is a rotation by an angle $\theta\left(  x\right)  $
around the $\mathbf{-\hat{x}}$ axis. The operator that affects the wave
function in this manner is \cite{rot}
\[
R=\exp\left[  -i\dfrac{\theta}{2}\hat{\sigma}_{x}\right]  =\cos\left(
\theta/2\right)  -i\hat{\sigma}_{x}\sin\left(  \theta/2\right)  ,
\]
while its inverse is given by
\[
R^{-1}=\exp\left[  i\dfrac{\theta}{2}\hat{\sigma}_{x}\right]  =\cos\left(
\theta/2\right)  +i\hat{\sigma}_{x}\sin\left(  \theta/2\right)  .
\]
It is easily verified that the transformation indeed diagonalizes the magnetic
part of the Hamiltonian,%

\[
RH_{M}R^{-1}=-\mu B\left(  x\right)  \hat{\sigma}_{z}.
\]
For the kinetic part we find
\[
RH_{K}R^{-1}=-\dfrac{\hbar^{2}}{2m}\left[  \dfrac{\partial^{2}}{\partial
x^{2}}-\dfrac{1}{4}\left(  \dfrac{d\theta}{dx}\right)  ^{2}+i\left(
\dfrac{d\theta}{dx}\dfrac{\partial}{\partial x}+\dfrac{1}{2}\dfrac{d^{2}%
\theta}{dx^{2}}\right)  \hat{\sigma}_{x}\right]  .
\]
Note that $i\left(  \dfrac{d\theta}{dx}\dfrac{\partial}{\partial x}+\dfrac
{1}{2}\dfrac{d^{2}\theta}{dx^{2}}\right)  \hat{\sigma}_{x}$ \emph{is} Hermitian.

Thus, the Hamiltonian of the system in the rotated frame may be written as
\begin{equation}
H=H_{diag}+H_{int}\label{h7}%
\end{equation}
where
\begin{align}
H_{diag} &  =-\dfrac{\hbar^{2}}{2m}\left[  \dfrac{\partial^{2}}{\partial
x^{2}}-\dfrac{1}{4}\left(  \dfrac{d\theta}{dx}\right)  ^{2}\right]  -\mu
B\left(  x\right)  \hat{\sigma}_{z}\label{h7.01}\\
H_{int} &  =-i\dfrac{\hbar^{2}}{2m}\left(  \dfrac{d\theta}{dx}\dfrac{\partial
}{\partial x}+\dfrac{1}{2}\dfrac{d^{2}\theta}{dx^{2}}\right)  \hat{\sigma}%
_{x}.\nonumber
\end{align}
The first part of the Hamiltonian $H_{diag}$ is diagonal. It contains the
kinetic part $\sim\partial^{2}/\partial x^{2}$, a term whose form is $\mp$
$\mu B(x)$ which is to be identified as the adiabatic effective potential, and
a term $\sim\left(  d\theta/dx\right)  ^{2}$ which appear due to the rotation.
The second part of the Hamiltonian $H_{int}$ contains only non-diagonal
components. These will be shown to be of order $\mathcal{O}\left(  K\right)  $
and hence may be regarded as a small perturbation. We proceed to find the
eigenstates of $H_{diag}$.

\subsection{Stationary states of $H_{diag}$.\label{diag}}

Since $H_{diag}$ is diagonal, the two spin states of the wavefunction are
decoupled. We then seek a solution of the form
\begin{equation}
\Psi_{\downarrow}=\left(
\begin{array}
[c]{c}%
0\\
\psi_{\downarrow}(x)
\end{array}
\right)  \text{ ; }E=E_{\downarrow}, \label{h8.01}%
\end{equation}
referred to as the \emph{spin-down} state, and another solution
\begin{equation}
\Psi_{\uparrow}=\left(
\begin{array}
[c]{c}%
\psi_{\uparrow}(x)\\
0
\end{array}
\right)  \text{ ; }E=E_{\uparrow}, \label{h8.02}%
\end{equation}
which we call the \emph{spin-up} state.

The equation for the non-vanishing component of the spin-down state is given
by
\begin{equation}
\left[  -\dfrac{\hbar^{2}}{2m}\left(  \dfrac{\partial^{2}}{\partial x^{2}%
}-\dfrac{1}{4}\left(  \dfrac{d\theta}{dx}\right)  ^{2}\right)  +\mu B\left(
x\right)  \right]  \psi_{\downarrow}=E_{\downarrow}\psi_{\downarrow},
\label{h8.1}%
\end{equation}
whereas the equation for the non-vanishing component of the spin-up state is
\begin{equation}
\left[  -\dfrac{\hbar^{2}}{2m}\left(  \dfrac{\partial^{2}}{\partial x^{2}%
}-\dfrac{1}{4}\left(  \dfrac{d\theta}{dx}\right)  ^{2}\right)  -\mu B\left(
x\right)  \right]  \psi_{\uparrow}=E_{\uparrow}\psi_{\uparrow}. \label{h8.2}%
\end{equation}

We now show that in the limit of small $K$ we can neglect the term
$\sim\left(  d\theta/dx\right)  ^{2}$ in both Eq.(\ref{h8.1}) and
Eq.(\ref{h8.2}): We compare the order of magnitude of the term $\mu B\left(
x\right)  $ to that of the term $\hbar^{2}\left(  d\theta/dx\right)  ^{2}/8m$.
Using Eq.(\ref{h1.1}) it can be easily shown that the maximum value of
$d\theta/dx$ is $B_{\bot}^{\prime}/B_{0}$ whereas the minimum value of $\mu
B\left(  x\right)  $ is $\mu B_{0}$. Thus,
\[
\dfrac{\left.  \mu B\left(  x\right)  \right|  _{\text{min}}}{\left(
\dfrac{\hbar^{2}}{8m}\left(  \dfrac{d\theta}{dx}\right)  _{\text{max}}%
^{2}\right)  }=\dfrac{8\mu mB_{0}^{3}}{\left(  B_{\bot}^{\prime}\right)
^{2}\hbar^{2}}=\frac{2}{K^{2}},
\]
and we reach the conclusion that when $K$ is small enough we can neglect the
term $\sim(d\theta/dx)^{2}$. Under this approximation, Eqs.(\ref{h8.1}) and
(\ref{h8.2}) simplify to
\begin{equation}
\left[  -\dfrac{\hbar^{2}}{2m}\dfrac{\partial^{2}}{\partial x^{2}}+\mu
B\left(  x\right)  \right]  \psi_{\downarrow}=E_{\downarrow}\psi_{\downarrow}
\label{h8.3}%
\end{equation}
and
\begin{equation}
\left[  -\dfrac{\hbar^{2}}{2m}\dfrac{\partial^{2}}{\partial x^{2}}-\mu
B\left(  x\right)  \right]  \psi_{\uparrow}=E_{\uparrow}\psi_{\uparrow}.
\label{h8.4}%
\end{equation}

The approximate solutions of these equations are outlined in the next two subsections.

\subsubsection{Stationary spin-down states.\label{down}}

Eq.(\ref{h8.3}) represents a particle in a symmetric \emph{attractive}
potential. If the extent of the wave function is small enough, we can expand
$B\left(  x\right)  $ to second order in $x$%
\begin{equation}
B\left(  x\right)  \simeq B_{0}\left[  1+\dfrac{1}{2}\left(  \dfrac{B_{\bot
}^{\prime}x}{B_{0}}\right)  ^{2}\right]  \label{down0}%
\end{equation}
and apply the well-known solution of a harmonic oscillator in one dimension.
We now derive the condition for which this approximation is valid: We recall
that the ground-state wave function of the harmonic oscillator is given
by\cite{sho}
\begin{equation}
\psi_{\downarrow}\left(  x\right)  =\left(  \dfrac{m\omega_{vib}}{\pi\hbar
}\right)  ^{1/4}e^{-\dfrac{2m\omega_{vib}x^{2}}{\hbar}}.\label{down1}%
\end{equation}
The extent of this wave function over which it changes appreciably is given
by
\begin{equation}
\Delta x_{\downarrow}\sim\sqrt{\dfrac{\hbar}{2m\omega_{vib}}},\label{down2}%
\end{equation}
whereas the extent over which $\mu B\left(  x\right)  $ changes significantly
(see Eq.(\ref{h1.1})) is
\begin{equation}
\Delta x_{\mu B}\sim B_{0}/B_{\bot}^{\prime}.\label{down3}%
\end{equation}
Thus, the ratio between these two length scales is
\begin{equation}
\dfrac{\Delta x_{\downarrow}}{\Delta x_{\mu B}}\sim\sqrt{\dfrac{\hbar\left(
B_{\bot}^{\prime}\right)  ^{2}}{2m\omega_{vib}^{2}B_{0}^{2}}}\sim\sqrt
{K}.\label{down4}%
\end{equation}
We therefore conclude that when $K$ is small enough, the harmonic
approximation is justified. The wave function $\psi_{\downarrow}\left(
x\right)  $, given by Eq.(\ref{down1}), then represents the lowest possible
bound state for this system. This state corresponds to a \emph{trapped}
particle. The energy of this state is clearly
\begin{equation}
E_{\downarrow}=\mu B_{0}+\frac{\hbar}{2}\omega_{vib}=\mu B_{0}\left(
1+K\right)  \simeq\mu B_{0},\label{down5}%
\end{equation}
while its full spinor representation is
\begin{equation}
\Psi_{\downarrow}=\left(
\begin{array}
[c]{c}%
0\\
\left(  \dfrac{m\omega_{vib}}{\pi\hbar}\right)  ^{1/4}e^{-\dfrac
{2m\omega_{vib}x^{2}}{\hbar}}%
\end{array}
\right)  .\label{down6}%
\end{equation}

\subsubsection{Stationary spin-up states.\label{up}}

Eq.(\ref{h8.4}) describes a particle in a \emph{repulsive} potential. It
corresponds to an unbounded state representing an \emph{untrapped} particle.
In this case there is a continuum of states, each with its own energy. As we
are interested in non-radiative decay, we focus on finding a solution with an
energy which is \emph{equal} to the energy found for the trapped state, that
is
\begin{equation}
E_{\uparrow}=E_{\downarrow}\simeq\mu B_{0}. \label{up0}%
\end{equation}
Since $H_{diag}$ is symmetric under $x\rightarrow-x$, to each energy in the
continuous spectrum there belongs one state with \emph{even} parity and one
state with \emph{odd} parity. Further, $H_{int}$ connects states with
\emph{opposite} parity only. Therefore, since the spin-down ground state is
\emph{even} in $x$, we need the spin-down state with \emph{odd }parity.

When evaluating the lifetime in the next section, we compute the matrix
element of $H_{int}$ between the states $\psi_{\uparrow}(x)$ and
$\psi_{\downarrow}(x)$. Thus, most of the contribution to this integral comes
from the region in $x$ where $\psi_{\downarrow}(x)$ is substantial. According
to Eq.(\ref{down4}), $\mu B(x)$ changes very little in this range and, as a
first approximation, we may take the potential in this region as
\emph{uniform},%
\begin{equation}
\mu B(x)\simeq\mu B_{0}\label{up0.1}%
\end{equation}
in Eq.(\ref{h8.4}), and then its solution becomes%
\begin{equation}
\psi_{\uparrow}(x)=C\sin\left(  \dfrac{\sqrt{4m\mu B_{0}}}{\hbar}x\right)
,\label{up0.2}%
\end{equation}
with the full spinor representation%
\begin{equation}
\Psi_{\uparrow}(x)=\left(
\begin{array}
[c]{c}%
C\sin\left(  \dfrac{\sqrt{4m\mu B_{0}}}{\hbar}x\right)  \\
0
\end{array}
\right)  .\label{up0.3}%
\end{equation}

The approximate wave function given in Eq.(\ref{up0.2}) is periodic near the
origin, and has a period of
\begin{equation}
\Delta x_{\uparrow}=\dfrac{\pi\hbar}{\sqrt{\mu mB_{0}}}.\label{up5}%
\end{equation}
Comparing it to $\Delta x_{\downarrow}$ given in Eq.(\ref{down2}), we find
that
\begin{equation}
\dfrac{\Delta x_{\uparrow}}{\Delta x_{\downarrow}}\sim\sqrt{K},\label{up6}%
\end{equation}
which shows that, when $K\ll1$, the wavefunction $\psi_{\uparrow}\left(
x\right)  $ executes many oscillations in the region where $\psi_{\downarrow
}\left(  x\right)  $ is appreciable.

\subsection{The lifetime.\label{time}}

To evaluate the lifetime $T_{esc}$ of the particle in its trapped state, which
is the average time it takes for the particle to escape, we calculate the
transition rate from the bound state given by Eq.(\ref{down6}), to the
unbounded state Eq.(\ref{up0.3}), according to Fermi's golden rule\cite{fermi}%
. Thus,
\begin{equation}
\dfrac{1}{T_{esc}}=\dfrac{2\pi}{\hbar}\left|  H_{\downarrow,\uparrow}\right|
^{2}g(E_{\uparrow}) \label{t1}%
\end{equation}
where
\begin{equation}
H_{\downarrow,\uparrow}=%
{\displaystyle\int\limits_{-\infty}^{+\infty}}
\Psi_{\downarrow}^{\dagger}H_{int}\Psi_{\uparrow}dx=%
{\displaystyle\int\limits_{-\infty}^{+\infty}}
dx\psi_{\downarrow}^{\ast}\left(  x\right)  \left(  \dfrac{\hbar^{2}}%
{2m}\right)  \left(  \dfrac{i}{2}\dfrac{d^{2}\theta}{dx^{2}}+i\dfrac{d\theta
}{dx}\dfrac{\partial}{\partial x}\right)  \psi_{\uparrow}\left(  x\right)
\label{t2}%
\end{equation}
is the matrix element of $H_{int}$ Eq.(\ref{h7.01}) between $\Psi_{\downarrow
}$ and $\Psi_{\uparrow}$, and $g(E_{\uparrow})$ is the density of the final
states at energy $E_{\uparrow}$.

Using Eq.(\ref{up0.2}) and Eq.(\ref{h1.1}) we find that%
\[
\dfrac{\dfrac{i}{2}\dfrac{d^{2}\theta}{dx^{2}}\psi_{\uparrow}}{\dfrac{d\theta
}{dx}\dfrac{\partial\psi_{\uparrow}}{\partial x}}\sim\dfrac{\left(
\dfrac{B_{\bot}^{\prime}}{B_{0}}\right)  ^{2}}{\dfrac{B_{\bot}^{\prime}}%
{B_{0}}\left(  \dfrac{\sqrt{4m\mu B_{0}}}{\hbar}\right)  }\sim K.
\]
Thus, when $K$ is small we may neglect the contribution of $d^{2}\theta
/dx^{2}$ to the integral Eq.(\ref{t2}) and then we find that
\begin{equation}
H_{\downarrow,\uparrow}=\dfrac{i\hbar^{2}}{2m}%
{\displaystyle\int\limits_{-\infty}^{+\infty}}
dx\psi_{\downarrow}^{\ast}\left(  x\right)  \dfrac{d\theta}{dx}\dfrac
{\partial\psi_{\uparrow}\left(  x\right)  }{\partial x}.\label{t4}%
\end{equation}

The integrand in Eq.(\ref{t4}) consists of a product of three functions: The
function $\psi_{\downarrow}^{\ast}$ whose `width' is about $\Delta
x_{\downarrow}$ (given in Eq.(\ref{down2})) around the origin, the function
$d\theta/dx$ whose extent around the origin $\Delta x_{\mu B}$ is roughly
$\sqrt{1/K}$ larger than $\Delta x_{\downarrow}$ and the function
$\partial\psi_{\uparrow}/\partial x$ which is a \emph{periodic} function with
a characteristic period $\Delta x_{\uparrow}$ which is $\sqrt{K}$
\emph{smaller} than $\Delta x_{\downarrow}$. This suggests that we can
approximate the integral in Eq.(\ref{t4}) by substituting $d\theta/dx$ for its
value at $x=0$,
\begin{equation}
\dfrac{d\theta}{dx}\simeq\dfrac{B_{\bot}^{\prime}}{B_{0}}. \label{t5}%
\end{equation}

Substituting Eqs.(\ref{t5}), (\ref{down1}) and (\ref{up0.2}) into
Eq.(\ref{t4}) gives
\begin{equation}
H_{\downarrow,\uparrow}\simeq i\dfrac{\hbar}{2m^{3/4}}C\sqrt{\dfrac
{2\mu\left(  B_{\bot}^{\prime}\right)  ^{2}}{B_{0}}}\left(  \dfrac{\pi\hbar
}{\omega_{vib}}\right)  ^{1/4}\exp\left[  -\dfrac{\mu B_{0}}{2\hbar
\omega_{vib}}\right]  , \label{t8}%
\end{equation}
where we have used the definite integral
\begin{equation}%
{\displaystyle\int\limits_{-\infty}^{+\infty}}
e^{-ax^{2}}\cos\left(  bx\right)  dx=\sqrt{\dfrac{\pi}{a}}\exp\left[
-\dfrac{b^{2}}{4a}\right]  . \label{t8.1}%
\end{equation}

When Eq.(\ref{t8}) is substituted into Eq.(\ref{t1}) the term $\left|
C\right|  ^{2}g(E_{\uparrow})$ appears. This term can be calculated by
temporarily introducing suitable boundary conditions: Assume that the system
is bounded by an infinite potential wall at $x=\pm L/2$, the length $L$ being
large compared to $\Delta x_{\downarrow}$ yet small when compared to $\Delta
x_{\mu B}$. In this case, the uniform potential approximation still holds, and
we may use the well-known result for the density of states for a particle in a
1D infinite potential well,%
\begin{equation}
g(E_{\uparrow}=\mu B_{0})=\sqrt{\dfrac{mL^{2}}{4\mu B_{0}\pi^{2}\hbar^{2}}%
}.\label{t8.2}%
\end{equation}
The evaluation of the normalization constant $C$ gives%
\begin{equation}
\left|  C\right|  ^{2}=2/L,\label{t8.3}%
\end{equation}
and therefore%
\begin{equation}
\left|  C\right|  ^{2}g(E_{\uparrow})=\sqrt{\dfrac{m}{\mu B_{0}\pi^{2}%
\hbar^{2}}}.\label{t13}%
\end{equation}

Finally, using Eqs.(\ref{t13}) and (\ref{t8}) inside Eq.(\ref{t1}) gives
\begin{equation}
T_{esc}=\dfrac{1}{\omega_{vib}}\sqrt{\dfrac{1}{2\pi K}}\exp\left[  \dfrac
{1}{2K}\right]  =T_{vib}\dfrac{1}{\sqrt{\left(  2\pi\right)  ^{3}K}}%
\exp\left[  \dfrac{1}{2K}\right]  , \label{t14}%
\end{equation}
where $T_{vib}=2\pi/\omega_{vib}$ is the period of classical oscillations
inside the trap.

Looking back at the calculation of the matrix element, Eq.(\ref{t8}), one
realizes that for small $K$ the wave function $\psi_{\uparrow}$ oscillates
strongly in the region where the wave function $\psi_{\downarrow}$ is
essentially different from zero. Thus, successive subintervals compensate each
other very effectively, and it is possible that the regions $\left|  x\right|
>\Delta x_{\downarrow}$ where the $x$-dependence of the potential becomes
important, contribute more effectively to the integral than the simple
estimates indicate. We have therefore carried out a more accurate calculation,
using the WKB approximation\cite{wkb1}, which takes the $x$-dependence of the
potential into account. The calculation, which is given in the Appendix,
yields the result
\begin{equation}
T_{esc}\simeq\dfrac{1}{\omega_{vib}}\dfrac{33}{32}\sqrt{\dfrac{1}{2\pi K}}%
\exp\left[  \dfrac{2}{K}\arctan\left(  \dfrac{1}{4}\right)  \right]
\label{t15}%
\end{equation}
which does indeed differ from Eq.(\ref{t14}) not only by the prefactor, being
unity in Eq.(\ref{t14}) and $33/32$ in Eq.(\ref{t15}), but even by the
exponent which is $0.5$ in Eq.(\ref{t14}) and $2\arctan\left(  1/4\right)
\simeq0.\,\allowbreak489\,96$ in Eq.(\ref{t15}).

\section{Discussion.\label{dis}}

Summarizing all we have found we conclude that the problem we have studied has
three important time scales: The shortest time scale is $T_{prec}$, which is
the time required for \emph{one} precession of the spin around the axis of the
local magnetic field. The intermediate time scale is $T_{vib}=T_{prec}/K$,
which is the time required to complete one cycle of the center of mass around
the center of the trap. These two time scales appear both in the classical and
the quantum-mechanical analysis. The longest time scale (provided $K$ is
small) $T_{esc}$, which is not present in the classical problem, is the time
it takes for the particle to escape from the trap.

Whereas the classical analysis yields an upper bound of $K=0.5$ for trapping
to occur, no such sharp bound exists in the quantum-mechanical analysis. This
is related to the fact that one cannot associate an effective potential well
with a finite barrier with the system. Nevertheless it is interesting to
compare the classical bound with the value of $K$ for which the exponent in
the expression for the quantum-mechanical lifetime becomes equal to $1$: For
the uniform-field approximation, Eq.(\ref{t14}), the two values happen to
agree exactly, and in the WKB approximation, Eq.(\ref{t15}), they differ only
by a small amount. Thus, the quantum-mechanical condition for trapping to
occur is essentially the same as the classical condition.

As an example, we apply our results to the case of a neutron and an atom
trapped with a field $B_{0}=100$ Oe and $B_{0}/B_{\bot}^{\prime}=10$cm. These
parameters correspond to typical traps used in Bose-Einstein condensation
experiments\cite{bec}. The results, being correct to within an order of
magnitude, are outlined in the following table:%

\[%
\begin{tabular}
[c]{|c||c|c|}\hline
& \multicolumn{2}{||c|}{$%
\begin{array}
[c]{c}%
B_{0}=100\text{ Oe}\\
B_{0}/B_{\bot}^{\prime}=10\text{cm}%
\end{array}
$}\\\cline{2-3}%
& Neutron & Atom\\\hline
$m$ gr & $\sim10^{-25}$ & $\sim10^{-22}$\\\hline
$\mu$ emu & $\sim10^{-23}$ & $\sim10^{-20}$\\\hline
$K$ & $\sim10^{-5}$ & $\sim10^{-8}$\\\hline
$T_{prec}$ sec & $\sim10^{-6}$ & $\sim10^{-9}$\\\hline
$T_{vib}$ sec & $\sim10^{-1}$ & $\sim10^{-1}$\\\hline
$T_{esc}$ sec & $\sim10^{\left(  10^{4}\right)  }$ & $\sim10^{\left(
10^{7}\right)  }$\\\hline
\end{tabular}
\]
We note that in both cases $K$ is very small compared to $1$. Consequently,
the calculated lifetime of the particle in the trap is extremely large,
suggesting that the particle (either neutron or atom) is tightly trapped in
this field.

Though the toy model presented in this paper is very simple, preliminary
studies\cite{tobe} show that behavior similar to what we found in this model
trap appears also in more realistic magnetic traps such as those used in
Bose-Einstein condensation experiments. For example, a similar analysis of a
two-dimensional Ioffee trap\cite{t2} shows similar behavior when the classical
analysis and quantum-mechanical analysis are compared, but in this case we
find that $T_{esc}\sim\exp\left[  2/K^{3}\right]  $.

The problem studied in this paper deals with a spin $1/2$ particle. Though
this fact has little influence on the solution of the \emph{classical}
problem, the extension to higher spin values complicates the analysis of the
\emph{quantum-mechanical} problem. In this case one has to deal with a
($2S+1$)-component spinor, and the interaction Hamiltonian does no longer
connect the ($-S$)-state to the ($+S$)-state, but only to the ($-S+1$) and
($-S+2$) states which for $S\geq5/2$ will \emph{still} be trapped.

\appendix

\section{Calculation of $\psi_{\uparrow}$ and $T_{esc}$ by the WKB
method.\label{app}}

In this appendix, we outline the solution of Eq.(\ref{h8.4}) by the WKB
approximation and calculate the resulting lifetime. The validity of the WKB
approximation is guaranteed by the following argument: We first note that, for
$E_{\uparrow}=\mu B_{0}$, the local de-Broglie wavelength $\lambda\left(
x\right)  $ is given by
\begin{equation}
\lambda\left(  x\right)  =\dfrac{2\pi\hbar}{\left.  p_{E_{\uparrow}%
}(x)\right|  _{@E_{\uparrow}=\mu B_{0}}}\label{app1}%
\end{equation}
where
\begin{equation}
p_{E_{\uparrow}}(x)=\sqrt{2m(E_{\uparrow}+\mu B\left(  x\right)  )}\simeq
\sqrt{2m\mu B_{0}\left(  2+\dfrac{1}{2}\left(  \dfrac{B_{\perp}^{\prime}%
x}{B_{0}}\right)  ^{2}\right)  }\label{app2}%
\end{equation}
is the classical momentum of the particle. Thus, the rate of change of
$\lambda$ over $x$ is
\begin{equation}
\left|  \dfrac{d\lambda}{dx}\right|  =2\pi\hbar\mu m\left[  2m(\mu B_{0}+\mu
B\left(  x\right)  \right]  ^{-3/2}\left|  \dfrac{\partial B(x)}{\partial
x}\right|  .\label{app3}%
\end{equation}
Its maximum value is given by
\begin{equation}
\left|  \dfrac{d\lambda}{dx}\right|  _{\text{max}}\sim\dfrac{\mu m\hbar
}{\left[  2m(\mu B_{0}+\mu B_{0})\right]  ^{3/2}}B^{\prime}\sim\sqrt
{\dfrac{\hbar^{2}\left(  B_{\bot}^{\prime}\right)  ^{2}}{\mu mB_{0}^{3}}}\sim
K.\label{app4}%
\end{equation}
Hence, in the adiabatic approximation, where $K\ll1$, the validity condition
of the WKB approximation $\left|  d\lambda/dx\right|  \ll1$ is satisfied.

In this approximation the wavefunction $\psi_{\uparrow}(x)$ corresponding to
the energy $E_{\uparrow}$ is given by\cite{wkb1}
\begin{equation}
\psi_{\uparrow}(x)=\dfrac{D}{\sqrt{\left|  p_{E_{\uparrow}}(x)\right|  }}%
\sin\left(  \dfrac{1}{\hbar}%
{\displaystyle\int\limits_{a}^{x}}
p_{E_{\uparrow}}(x^{\prime})dx^{\prime}\right)  , \label{app5}%
\end{equation}
where $a$, the lower limit of the integration, is yet undetermined.

We still need to evaluate the product $\left|  D\right|  ^{2}g(E_{\uparrow
}=\mu B)$ which appears when calculating $T_{esc}$ by Eq.(\ref{t1}). We
therefore temporarily introduce boundary conditions by assuming that the
system is bounded by an infinite potential wall at $x=\pm L/2$. Thus, by
setting $a=-L/2$ the boundary condition at $x=-L/2$ is automatically
satisfied. We note that we are looking for a \emph{highly excited} state
$\psi_{\uparrow}$ and we consider the eigenstate corresponding to the nearest
higher energy $E_{\uparrow}+\Delta E$. The latter must fulfill the requirement
that the \emph{phase} of $\psi$ at $x=L/2$ changes by $\pi$ when going from
$E_{\uparrow}$ to $E_{\uparrow}+\Delta E$. Thus,
\begin{equation}
\dfrac{1}{\hbar}\left[  \int_{-L/2}^{L/2}p_{E_{\uparrow}+\Delta E}(x^{\prime
})dx^{\prime}-\int_{-L/2}^{L/2}p_{E_{\uparrow}}(x^{\prime})dx^{\prime}\right]
=\pi.\label{app6}%
\end{equation}
Since $E_{\uparrow}\gg\mu B(x)$ everywhere we may write the difference in
Eq.(\ref{app6}) as a derivative
\begin{equation}
\dfrac{1}{\hbar}\Delta E\dfrac{\partial}{\partial E_{\uparrow}}\int
_{-L/2}^{L/2}p_{E_{\uparrow}}(x^{\prime})dx^{\prime}=\pi.\label{app7}%
\end{equation}
Using Eq.(\ref{app2}) we rewrite Eq.(\ref{app7}) as
\begin{equation}
\dfrac{1}{\hbar}\Delta E\int_{-L/2}^{L/2}\dfrac{m}{p_{E_{\uparrow}}(x^{\prime
})}dx^{\prime}=\pi.\label{app8}%
\end{equation}
Normalization of the wavefunction Eq.(\ref{app5}), gives on the other-hand
\begin{align}
\int_{-L/2}^{L/2}\left|  \psi_{\uparrow}(x)\right|  ^{2}dx &  \simeq\left|
D\right|  ^{2}\int_{-L/2}^{L/2}\dfrac{1}{\left|  p_{E_{\uparrow}}(x)\right|
}\sin^{2}\left(  \dfrac{1}{\hbar}%
{\displaystyle\int\limits_{-L/2}^{x}}
p_{E_{\uparrow}}(x^{\prime})dx^{\prime}\right)  dx\label{app9}\\
&  \simeq\dfrac{1}{2}\left|  D\right|  ^{2}\int_{-L/2}^{L/2}\dfrac{1}%
{p_{E}(x)}dx=1,\nonumber
\end{align}
where we have neglected the terms containing fast oscillations by replacing
the $\sin^{2}()$ term by its average value $1/2$ . Substituting Eq.(\ref{app9}%
) into Eq.(\ref{app8}) gives
\begin{equation}
\left|  D\right|  ^{2}\dfrac{1}{\Delta E}=\left|  D\right|  ^{2}g(E_{\uparrow
})=\dfrac{2m}{\pi\hbar}\label{app10}%
\end{equation}
where $g(E_{\uparrow})=1/\Delta E$ is the density of states at energy
$E_{\uparrow}$.

To evaluate the integral in Eq.(\ref{t4}) we use the \emph{stationary phase
}method\cite{phase}. Using Eqs.(\ref{app5}), (\ref{down1}) and (\ref{t5}) we
rewrite this integral as
\[
H_{\uparrow,\downarrow}\simeq i\dfrac{\hbar^{2}}{2m}\int\psi_{\uparrow}^{\ast
}\left(  x\right)  \dfrac{d\theta}{dx}\dfrac{\partial\psi_{\downarrow}%
}{\partial x}dx=I_{+}-I_{-}%
\]
where we have defined
\begin{equation}
I_{\pm}=A%
{\displaystyle\int\limits_{-\infty}^{+\infty}}
dxg(x)\exp f_{\pm}(x),\nonumber
\end{equation}%
\[
A\equiv\dfrac{\hbar^{2}}{4m}\left(  \dfrac{m\omega_{vib}}{2\pi\hbar}\right)
^{1/4}\left(  \dfrac{4m\omega_{vib}}{\hbar}\right)  D\left(  \dfrac{B_{\bot
}^{\prime}}{B_{0}}\right)  ,
\]%
\[
g(x)\equiv\dfrac{x}{\sqrt{\left|  p_{E_{\uparrow}}\left(  x\right)  \right|
}},
\]
and
\begin{equation}
f_{\pm}(x)\equiv-\dfrac{2m\omega_{vib}x^{2}}{\hbar}\pm\dfrac{i}{\hbar}%
{\displaystyle\int\limits_{0}^{x}}
p_{E_{\uparrow}}\left(  x^{\prime}\right)  dx^{\prime}. \label{fpm}%
\end{equation}
Note that the lower integration limit in $f_{\pm}\left(  x\right)  $ has been
chosen as $0$ to make the function $\psi_{\uparrow}^{\ast}\left(  x\right)  $
antisymmetric under $x\rightarrow-x$.

We continue to work out $I_{+}$ first: According to the stationary phase
approximation we should first find the point $x_{s}$ for which the phase
$f_{+}(x)$ is stationary. This is accomplished by solving
\begin{equation}
\dfrac{\partial f_{+}}{\partial x}=\dfrac{\partial}{\partial x}\left[
-\dfrac{2m\omega_{vib}x^{2}}{\hbar}+\dfrac{i}{\hbar}%
{\displaystyle\int\limits^{x}}
p_{E_{\uparrow}}\left(  x^{\prime}\right)  dx^{\prime}\right]  =0\label{eqw.3}%
\end{equation}
for $x$. Using Eqs.(\ref{up0}) and (\ref{down0}) and after a few algebraic
steps we find that the only solution to Eq.(\ref{eqw.3}) is
\begin{equation}
x_{s}^{+}=i\sqrt{\dfrac{4}{17}}\dfrac{B_{0}}{B_{\bot}^{\prime}}.\label{eqw.6}%
\end{equation}
Now we approximate the integral $I_{+}$ as
\begin{equation}
I_{+}=A%
{\displaystyle\int\limits_{-\infty}^{+\infty}}
dxg(x)\exp\left[  f_{+}(x)\right]  \simeq Ag(x_{s}^{+})\sqrt{\dfrac{2\pi
}{-f_{+}^{\prime\prime}(x_{s}^{+})}}\exp\left[  f_{+}(x_{s}^{+})\right]
.\label{eqw.7}%
\end{equation}
The evaluation of $f_{+}(x_{s}^{+})$ is performed in the following way:
\begin{align*}
f_{+}(x_{s}^{+}) &  =-\dfrac{2m\omega_{vib}\left(  x_{s}^{+}\right)  ^{2}%
}{\hbar}+\dfrac{i}{\hbar}%
{\displaystyle\int\limits_{0}^{i\sqrt{4/17}B_{0}/B_{\perp}^{\prime}}}
p_{E_{\uparrow}}\left(  x\right)  dx\\
&  =\dfrac{2m}{\hbar}\sqrt{\dfrac{\left(  B_{\perp}^{\prime}\right)  ^{2}\mu
}{mB_{0}}}\left(  \sqrt{\dfrac{4}{17}}\dfrac{B_{0}}{B_{\perp}^{\prime}%
}\right)  ^{2}+\dfrac{i^{2}}{\hbar}%
{\displaystyle\int\limits_{0}^{\sqrt{4/17}B_{0}/B_{\perp}^{\prime}}}
p_{E_{\uparrow}}\left(  iy\right)  dy\\
&  =\dfrac{2m}{\hbar}\sqrt{\dfrac{\left(  B_{\perp}^{\prime}\right)  ^{2}\mu
}{mB_{0}}}\left(  \sqrt{\dfrac{4}{17}}\dfrac{B_{0}}{B_{\perp}^{\prime}%
}\right)  ^{2}-\dfrac{1}{\hbar}%
{\displaystyle\int\limits_{0}^{\sqrt{4/17}B_{0}/B_{\perp}^{\prime}}}
\sqrt{2m\mu B_{0}\left(  2-\dfrac{1}{2}\left(  \dfrac{B_{\perp}^{\prime}%
y}{B_{0}}\right)  ^{2}\right)  }dy\\
&  =-\sqrt{\dfrac{4\mu mB_{0}^{3}}{\left(  B_{\perp}^{\prime}\right)
^{2}\hbar^{2}}}\arcsin\left(  \frac{1}{\sqrt{17}}\right)  =-\dfrac
{\arctan\left(  1/4\right)  }{K}%
\end{align*}
where we used the definite integral%
\[
\int_{0}^{y}\sqrt{a^{2}-x^{2}}dx=\dfrac{1}{2}y\sqrt{a^{2}-y^{2}}+\dfrac{1}%
{2}a^{2}\arcsin\left(  \dfrac{y}{a}\right)  \text{ ; for }y<a\text{.}%
\]
The term $f_{+}^{\prime\prime}(x_{s}^{+})$ is calculated straightforward with
the result that
\begin{align}
f_{+}^{\prime\prime}(x_{s}^{+}) &  =\sqrt{\dfrac{m\mu\left(  B_{\bot}^{\prime
}\right)  ^{2}}{B_{0}}}\left(  -2-i\dfrac{1}{17\sqrt{1-1/17}}\right)
\label{eqw.8}\\
&  =-2\sqrt{\dfrac{m\mu\left(  B_{\bot}^{\prime}\right)  ^{2}}{B_{0}}}%
\sqrt{\dfrac{1089}{1088}}\exp\left[  i\arctan\left(  \dfrac{\sqrt{17}}%
{136}\right)  \right]  .\nonumber
\end{align}
Substituting this together with $g(x_{s}^{+})$ and $f(x_{s}^{+})$ into
Eq.(\ref{eqw.7}) gives
\begin{align}
I_{+} &  =iA\sqrt{\dfrac{B_{0}}{B_{\bot}^{\prime}}}\dfrac{\left(  17\right)
^{1/4}}{2}\sqrt{\dfrac{2\pi}{17}}\sqrt{\dfrac{B_{0}}{m\mu\left(  B_{\bot
}^{\prime}\right)  ^{2}}}\label{eqw.9}\\
&  \times\left(  \dfrac{1088}{1089}\right)  ^{1/4}\exp\left[  -\dfrac{1}%
{K}\arctan\left(  \dfrac{1}{4}\right)  -\dfrac{i}{2}\arctan\left(
\dfrac{\sqrt{17}}{136}\right)  \right]  .\nonumber
\end{align}
Likewise, a similar calculation for $I_{-}$ gives%
\begin{align*}
x_{s}^{-} &  =-x_{s}^{+}\\
f_{-}(x_{s}^{-}) &  =f_{+}(x_{s}^{+})\\
f_{-}^{\prime\prime}(x_{s}^{-}) &  =f_{+}^{\prime\prime}(x_{s}^{+})\\
g(x_{s}^{-}) &  =-g(x_{s}^{+}),
\end{align*}
leading to the result that
\begin{align}
I_{-} &  =-iA\sqrt{\dfrac{B_{0}}{B_{\bot}^{\prime}}}\dfrac{\left(  17\right)
^{1/4}}{2}\sqrt{\dfrac{2\pi}{17}}\sqrt{\dfrac{B_{0}}{m\mu\left(  B_{\bot
}^{\prime}\right)  ^{2}}}\label{eqw.10}\\
&  \times\left(  \dfrac{1088}{1089}\right)  ^{1/4}\exp\left[  -\dfrac{1}%
{K}\arctan\left(  \dfrac{1}{4}\right)  -\dfrac{i}{2}\arctan\left(
\dfrac{\sqrt{17}}{136}\right)  \right]  .\nonumber
\end{align}
By repeating the steps that led to Eq.(\ref{t14}) we find that
\begin{equation}
T_{esc}=\dfrac{1}{\omega_{vib}}\dfrac{33}{32}\sqrt{\dfrac{1}{2\pi K}}%
\exp\left[  \dfrac{2}{K}\arctan\left(  \dfrac{1}{4}\right)  \right]
.\label{eqw.14}%
\end{equation}
\end{document}